\documentclass[journal]{IEEEtran}

\usepackage{amsmath,epsfig,bm,amssymb,amsthm}
\usepackage{psfrag}
\usepackage{cite}
\usepackage{color}

\ifCLASSINFOpdf
\else
\fi

\usepackage{amsmath,epsfig,bm,amssymb,amsthm,balance}

\newcommand{\dd}{{\mathrm d}}

\newcommand{\CN}{\mathcal{CN}}
\newcommand{\Et}{\mathrm{E}}
\newcommand{\bt}{\mathrm{b}}
\newcommand{\Var}{\mathrm{Var}}
\newcommand{\opt}{\mathrm{opt}}
\newcommand{\lpn}{\lim \limits_{P_0/N_0 \rightarrow \infty}}

\begin{document}
%
\title{Performance of Selection Combining for Differential Amplify-and-Forward Relaying Over Time-Varying Channels
\thanks{The authors are with the Department of Electrical and Computer Engineering,
University of Saskatchewan, Saskatoon, Canada, S7N5A9.
Email: m.avendi@usask.ca, ha.nguyen@usask.ca.}}

\author{\IEEEauthorblockN{M. R. Avendi and Ha H. Nguyen}
}

\maketitle

\begin{abstract}
\label{abs}
Selection combining (SC) at the destination for differential amplify-and-forward (AF) relaying is attractive as it does not require channel state information as compared to the semi maximum-ratio-combining (semi-MRC) while delivering close performance. Performance analysis of the SC scheme was recently reported but only for the case of slow-fading channels. This paper provides an exact average bit-error-rate (BER) of the SC scheme over a general case of time-varying Rayleigh fading channels and when the DBPSK modulation is used together with the non-coherent detection at the destination. The presented analysis is thoroughly verified with simulation results in various fading scenarios. It is shown that the performance of the system is related to the auto-correlation values of the channels. It is also shown that the performance of the SC method is very close to that of the semi-MRC method and the existence of an error floor at high signal-to-noise ratio region is inevitable in both methods. The obtained BER analysis for the SC method can also be used to approximate the BER performance of the MRC method, whose exact analytical evaluation in time-varying channels appears to be difficult.
\end{abstract}


\IEEEpeerreviewmaketitle

\section{Introduction}
\label{se:intro}

Cooperative communications has now become a mature research topic. Currently, a special type of cooperative communications (with the help of one relay) has been standardized in the 3 GPP LTE technology to leverage the coverage problem of cellular networks and it is envisaged that LTE-advanced version will include cooperative relay features to overcome other limitations such as capacity and interference \cite{coop-dohler}. There are also applications for cooperative relay networks in wireless LAN, vehicle-to-vehicle communications and wireless sensor networks that have been discussed in \cite{coop-LTE,coop-WiMAX,coop-deploy,coop-sensor} and references therein.

In cooperative communications, a user in the network acts as a relay to receive signals from a source, processes and re-broadcasts to a destination. In this way, additional links, other than the direct link from a source to a destination, can be constructed via relays and hence the overall spatial diversity of the system would be increased. Depending on the signal processing strategy that a relay utilizes, relay networks are generally classified as decode-and-forward (DF) and amplify-and-forward (AF) \cite{coop-laneman}.

Among these two strategies, AF or its non-coherent version, differential AF (D-AF) is very attractive as it requires less computational burden at the relays and destination. In D-AF, data symbols are differentially encoded at the source. The relay's function is simply to multiply the received signal with a fixed amplification factor. At the destination, the received signals from multi-links are combined to achieve the diversity, and used for non-coherent detection of the transmitted signals without the need of instantaneous channel state information (CSI). In \cite{DAF-Liu,DAF-DDF-QZ,DAF-General}, a maximum-ratio combiner using a set of fixed weights, based on the second-order statistics of all channels, has been used to combine the received signals from the relay-destination and source-destination links. For future reference, this combiner is called semi-maximum ratio combiner (semi-MRC).

With the motivation of reducing the detection complexity at the destination, selection combining for differential AF relay networks was recently investigated and analyzed in \cite{DAF-SC}. This combiner can be seen as a counterpart of selection combining of DPSK in point-to-point communications with receive diversity studied in \cite{SC-Kam1,SC-Norman,SC-Kam2,SC-DPSK}. However, the analysis reported in our previous work \cite{DAF-SC} only applies for \emph{slow-fading} and symmetric channels. The slow-fading assumption requires approximate equality of two consecutive channel uses, which would be violated in practice under high mobility of users. Moreover, the channels are not necessarily symmetric in practice, i.e., different channels might need to be characterized with different variances.

This paper studies D-AF relaying over \emph{time-varying} Rayleigh-fading channels {with non-identical (general) channel variances} using post-detection selection combining (SC). The DBPSK modulation is used and the AF strategy with fixed gain at the relay is employed. Two links are involved in the communication: the direct link from the source to the destination (SD) and the cascaded link from the source to the destination via the relay. The decision variable is computed for each link and the one with the maximum magnitude is chosen for non-coherent detection. Hence, different from the semi-MRC, the selection combiner does not need the second-order statistic of any of the channels, which simplifies the detection at the destination. To characterize the time-varying nature of the channels, first-order auto-regressive models \cite{AR1-ch,DAF-ITVT} are employed for the direct and cascaded channels. The probability density function (pdf) and cumulative density function (cdf) of the decision variable in each link are derived and used to obtain the exact average bit-error-rate (BER). The analysis is verified with simulation results in different fading and channel scenarios. Comparison of the SC and semi-MRC systems shows that the performance of the SC method is very close to that of the semi-MRC. For fast-fading channels, it is seen that the performance of both SC and semi-MRC systems degrades and reaches an error floor. The expression of the error floor is also derived for the SC method. On the other hand, the close performance of both the SC and semi-MRC schemes implies that one can use the exact BER analysis of the SC method to closely approximate the performance of the semi-MRC method in time-varying channels. This is useful since the exact BER of the semi-MRC method in time-varying channels appears to be difficult \cite{DAF-ITVT} and only a loose lower bound was derived for this system in \cite{DAF-ITVT}.

The outline of the paper is as follows. Section \ref{sec:system} describes the system model. In Section III the non-coherent detection of D-AF relaying using SC technique is developed. The performance of the system is considered in Section \ref{sec:symbol_error_probability}. {Section~\ref{sec:mult-relay} discusses extension to multi-relay systems.} Simulation results are given in Section \ref{sec:sim}. Section \ref{sec:con} concludes the paper.

\emph{Notation}: $(\cdot)^*$, $|\cdot|$ denote conjugate and absolute values of a complex number, respectively. $\mathcal{CN}(0,\sigma^2)$ stands for a complex Gaussian distribution with mean zero and variance $\sigma^2$, while $\chi_2^2$ stands for chi-squared
distribution with two degrees of freedom. $\Et\{\cdot\}$ and $\Var\{\cdot\}$ are expectation and variance operations, respectively. Both $\exp(\cdot)$ and ${\mathrm{e}}^{(\cdot)}$ indicate exponential function and $E_1(x)=\int \limits_{x}^{\infty} ({\mathrm{e}}^{-t}/t)\dd t$ is the exponential integral function.

\section{System Model}
\label{sec:system}
\begin{figure}[t]
\psfrag {Source} [] [] [1.0] {Source}
\psfrag {Relay} [] [] [1.0] {Relay}
\psfrag {Destination} [] [] [1.0] {Destination}
\psfrag {h1} [] [] [1.0] {$h_1[k]$\;\;\;}
\psfrag {h2} [] [] [1.0] {\;\;$h_2[k]$}
\psfrag {h0} [] [] [1.0] {\;\;$h_0[k]$}
\centerline{\epsfig{figure={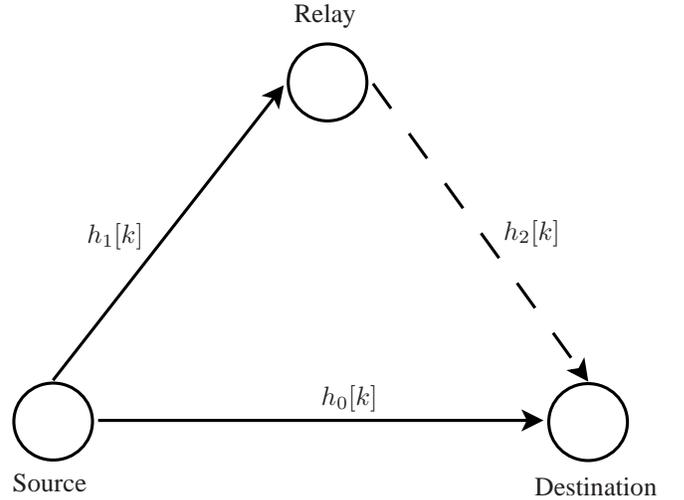},width=8.5cm}}
\caption{The wireless relay model under consideration.}
\label{fig:sysmodel}
\end{figure}

The system model in this article is very similar to that of \cite{DAF-WCNC,DAF-SC,DAF-ITVT}. As such, the formulation and description of the system model are similar to those in \cite{DAF-WCNC,DAF-SC,DAF-ITVT}. Figure~\ref{fig:sysmodel} depicts the wireless relay model under consideration, which has three nodes: one Source, one Relay and one Destination. There are a direct link and a cascaded link, via Relay, from Source to Destination. The inherent diversity order of the system is therefore two. A common half-duplex communication between the nodes is assumed, i.e., each node employs a single antenna and able to only send or receive in any given time.

The channel coefficients at time $k$, from Source to Destination (SD), from Source to Relay (SR) and from Relay to Destination (RD) are shown with $h_0[k]$, $h_1[k]$ and $h_2[k]$, respectively. A Rayleigh flat-fading model is assumed for each channel, i.e., $h_i\sim \CN(0,\sigma_i^2),\; i=0,1,2$. {If $\sigma_1^2=\sigma_2^2=\sigma_3^2,$ the channels are called \emph{symmetric}, otherwise they are called \emph{non-symmetric}}. The channels are spatially uncorrelated and changing continuously in time {(i.e., time-varying)}. The time correlation between two channel coefficients, $n$ symbols apart, follows the Jakes' model \cite{microwave-jake}:
\begin{equation}
\label{eq:phi}
\varphi_i(n)=\Et\{h_i[k] h_i^*[k+n]\}=\sigma_i^2 J_0(2\pi f_i n), \quad i=0,1,2
\end{equation}
where $J_0(\cdot)$ is the zeroth-order Bessel function of the first kind and $f_i$ is the maximum normalized Doppler frequency of the $i$th channel. 
{The amount of channel variation is related to the normalized Doppler shift which is a function of user's mobility and signalling rate. Hence, the normalized Doppler shift is used to distinguish between slow time-varying (slow-fading) and rapid time-varying (fast-fading) channels.}

Let $\mathcal{V}=\{-1,+1\}$ be the set of BPSK symbols. Each information bit at time $k$ is transformed to a BPSK symbol $v[k]\in \mathcal{V}$. Before transmission, the symbols are encoded differentially as
\begin{equation}
\label{eq:s-source}
s[k]=v[k] s[k-1],\quad s[0]=1.
\end{equation}

The transmission process is divided into two phases. A symbol or a frame of symbols could be transmitted in each phase. Symbol-by-symbol transmission is not practical as it causes frequent switching between reception and transmission. Hence, frame-by-frame transmission protocol is utilized here. However, the analysis is the same for both cases and only the channels auto-correlation values are different. In symbol-by-symbol transmission, two channel uses are two symbols apart $(n=2),$ while in frame-by-frame transmission two channel uses are one symbol apart $(n=1)$.

In the first phase, symbol $s[k]$ is transmitted by Source to Relay and Destination. Let $P_0$ be the average power of Source per symbol. The received signals at Destination and Relay are
\begin{equation}
\label{eq:y0}
y_0[k]=\sqrt{P_0}h_0[k]s[k]+w_0[k],
\end{equation}
\begin{equation}
\label{eq:relay_rx}
y_1[k]=\sqrt{P_0}h_1[k]s[k]+w_1[k],
\end{equation}
where $w_0[k],\; w_1[k]\sim \CN(0,N_0)$ are noise components at Destination and Relay, respectively. It is easy to see that, for given $s[k]$, $y_0[k]\sim \CN(0,N_0(\rho_0+1))$, where $\rho_0$ is the average received SNR per symbol from the direct link, defined as
\begin{equation}
\label{eq:rho0}
\rho_0=\frac{P_0 \sigma_0^2}{N_0}.
\end{equation}
Also, the average received SNR per symbol at Relay is defined as
\begin{equation}
\label{eq:rho_1}
\rho_1=\frac{P_0\sigma_1^2}{N_0}.
\end{equation}
The received signal at Relay is then multiplied by an amplification factor, and re-transmitted to Destination. Based on the variance of SR channel, the amplification factor commonly used in the literature is
\begin{equation}
\label{eq:A}
A=\sqrt{\frac{P_1}{P_0 \sigma_1^2+N_0}},
\end{equation}
where $P_1$ is the average transmitted power per symbol at Relay. In general, $A$ can be any arbitrarily fixed value. If the total power in the network, $P$, is divided between Source and Relay such that $P_0=qP,\;P_1=(1-q)P$, where $q$ is the power amplification factor, then $A=\sqrt{(1-q)P/(qP\sigma_1^2+N_0)}$.

The corresponding received signal at Destination is
\begin{equation}
\label{eq:y}
y_2[k]=A \; h_2[k]y_1[k]+w_2[k],
\end{equation}
where $w_2[k]\sim \mathcal{CN}(0,N_0)$ is the noise component at Destination in the second phase. Substituting (\ref{eq:relay_rx}) into (\ref{eq:y}) yields
\begin{equation}
\label{eq:destination-rx}
y_2[k]= A\; \sqrt{P_0}h[k]s[k]+w[k],
\end{equation}
where $h[k]=h_1[k]h_2[k]$ is the equivalent double-Rayleigh channel with zero mean and variance $\sigma_1^2 \sigma_2^2$ \cite{SPAF-P} and $w[k]=A\; h_2[k]w_1[k]+w_2[k]$ is the equivalent noise component. It should be noted that for a given $h_2[k]$, $w[k]$ is a complex Gaussian random variable with zero mean and variance
\begin{equation}
\label{eq:sig2wk}
\sigma_{w}^2=N_0(1+A^2 \; |h_2[k]|^2)
\end{equation}
and hence $y_2[k]$, conditioned on $s[k]$ and $h_2[k]$, is a complex Gaussian random variable with zero mean and variance $(\rho_2+1)\sigma_{w}^2$. Here, $\rho_2$ is the average received SNR per symbol from the cascaded link at Destination, conditioned on $h_2[k]$. It is given as
\begin{equation}
\label{eq:rhok}
\rho_2= \frac{A^2 \rho_1 |h_2[k]|^2}{1+A^2|h_2[k]|^2}.
\end{equation}

The next section presents the selection combining of the received signals at Destination and its non-coherent detection.

\section{Selection Combining and Non-Coherent Detection}
\label{sec:combinig}
Based on two consecutive received symbols, non-coherent detection of the transmitted symbols can be obtained. For DBPSK, the decision variables for the direct and cascaded links are computed from the two latest symbols as
\begin{align}
\label{eq:zeta_0}
\zeta_0= \Re\{y_0^*[k-1] y_0[k] \}, \\
\label{eq:zeta_1}
\zeta_2= \Re\{ y_2^*[k-1] y_2[k] \}.
\end{align}

To achieve the cooperative diversity, the decision variables from the two transmission phases should be combined using some combining technique \cite{Linear-Diversity}. For the semi-MRC method, over slow-fading channels, the decision variables were combined as \cite{DAF-Liu,DAF-DDF-QZ,DAF-General}
\begin{equation}
\label{eq:zeta}
\zeta=\frac{1}{N_0} \zeta_0+\frac{1}{N_0(1+A^2\sigma_2^2)} \zeta_2.
\end{equation}

\begin{figure}[t]
\psfrag {y1} [] [] [1.0] {$y_0[k]$}
\psfrag {y2} [] [] [1.0] {$y_2[k]$}
\psfrag {Delay} [] [] [1.0] {Delay}
\psfrag {Decision} [] [] [1.0] {Selection}
\psfrag {y1k} [l] [] [1.0] {$y_0^*[k-1]$}
\psfrag {y2k} [l] [] [1.0] {$y_2^*[k-1]$}
\psfrag {zeta1} [l] [] [1.0] {$\zeta_0$}
\psfrag {zeta2} [l] [] [1.0] {$\zeta_2$}
\psfrag {zeta} [l] [] [1.0] {$\zeta$}
\psfrag {*} [] [] [1.0] {*}
\centerline{\epsfig{figure={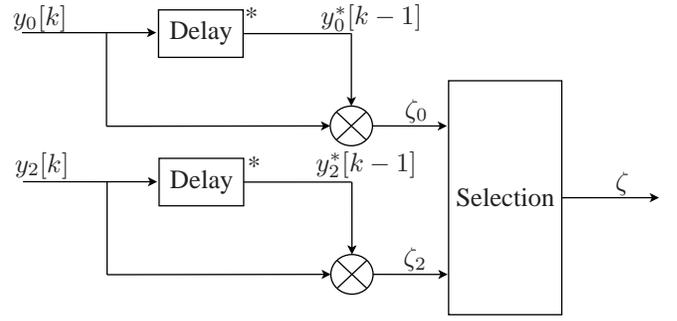},width=8.5cm}}
\caption{Block diagram of the selection combiner at Destination of a D-AF relay network.}
\label{fig:sc-block}
\end{figure}

However, instead of the semi-MRC which needs the second-order statistics of all channels, it is proposed to combine the received signals using a selection combiner as illustrated in Fig.~\ref{fig:sc-block} \cite{DAF-SC}. As it is seen, the decision statistics for the direct link, $\zeta_0$, and the cascaded link, $\zeta_2$, are computed and compared to choose the link with a higher magnitude. The output of the combiner is therefore
\begin{equation}
\label{eq:zeta-sc}
\zeta =
\begin{cases}
\zeta_0, & \mbox{if} \;\; |\zeta_0|>|\zeta_2|\\
\zeta_2, & \mbox{if} \;\; |\zeta_2|>|\zeta_0|.
\end{cases}
\end{equation}
Obviously, using this scheme, no channel information is needed at Destination.

Finally, the output of the combiner is used to decode the transmitted signal as
\begin{equation}
\label{eq:ml-detection}
\hat{v}[k]=
\begin{cases}
-1, & \mbox{if} \;\; \zeta<0\\
+1, & \mbox{if} \;\; \zeta>0
\end{cases}.
\end{equation}

The next section analyzes the performance of the above selection combining detector.

\section{Error Performance Analysis}
\label{sec:symbol_error_probability}
As usual, the transmitted symbols are assumed to be equally likely. Without loss of generality, assume that symbol $v[k]=+1$ is transmitted and let $\hat{v}[k]$ denote the decoded symbol. The BER can be expressed as
\begin{multline}
\label{eq:PbE}
P_{\bt}(E)=\Pr(\zeta<0, v[k]=+1)=\Pr(\zeta_0<0,|\zeta_0|>|\zeta_2|)\\
+\Pr(\zeta_2<0,|\zeta_2|>|\zeta_0|).
\end{multline}
Since $\zeta_0$ and $\zeta_2$ have different distributions, the two terms in \eqref{eq:PbE} should be computed separately. The first term can be written as
\begin{multline}
\label{eq:Pb1}
P_{\bt}(E_1)=\Pr(\zeta_0<0,|\zeta_0|>|\zeta_2|)\\
=\Pr(|\zeta_2|+\zeta_0<0)
= \int \limits_{-\infty}^{0} \int \limits_{0}^{-\beta} f_{\zeta_0}(\beta) f_{|\zeta_2|}(r) \dd r \dd \beta\\
=\int \limits_{-\infty}^{0} f_{\zeta_0}(\beta) \left[F_{|\zeta_2|}(-\beta)-F_{|\zeta_2|}(0)\right] \dd \beta.
\end{multline}
Likewise, the second term of \eqref{eq:PbE} can be expressed as
\begin{multline}
\label{eq:Pb2}
P_{\bt}(E_2)=\Pr(\zeta_2<0,|\zeta_2|>|\zeta_0|)\\
=\Pr(|\zeta_0|+\zeta_2<0)
= \int \limits_{-\infty}^{0} \int \limits_{0}^{-\beta} f_{\zeta_2}(\beta) f_{|\zeta_0|}(r) \dd r \dd \beta\\
=\int \limits_{-\infty}^{0} f_{\zeta_2}(\beta) \left[F_{|\zeta_0|}(-\beta)-F_{|\zeta_0|}(0)\right] \dd \beta.
\end{multline}

In \eqref{eq:Pb1} and \eqref{eq:Pb2}, $f_{\zeta_0}(\cdot)$ and $f_{\zeta_2}(\cdot)$ are the pdfs of $\zeta_0$ and $\zeta_2$, respectively. Also, $F_{|\zeta_0|}(\cdot)$ and $F_{|\zeta_2|}(\cdot)$ are the cdfs of $|\zeta_0|$ and $|\zeta_2|$, respectively. They can be written as
\begin{equation}
\label{eq:F_z1}
\begin{split}
F_{|\zeta_2|}(\beta)=\Pr(|\zeta_2|\leq \beta)=\Pr(-\beta\leq \zeta_2 \leq \beta)\\
= F_{\zeta_2}(\beta)-F_{\zeta_2}(-\beta),
\end{split}
\end{equation}
\begin{equation}
\label{eq:F_|z0|}
\begin{split}
F_{|\zeta_0|}(\beta)=\Pr(|\zeta_0|\leq \beta)=\Pr(-\beta \leq \zeta_0 \leq \beta)\\
= F_{\zeta_0}(\beta)-F_{\zeta_0}(-\beta),
\end{split}
\end{equation}
where $F_{\zeta_0}(\cdot)$ and $F_{\zeta_2}(\cdot)$ are the cdfs of $\zeta_0$ and $\zeta_2$, respectively.

To proceed with the computation of \eqref{eq:Pb1} and \eqref{eq:Pb2}, the pdfs and cdfs of $\zeta_0$ and $\zeta_2$ are required. To obtain these functions, the relationship between two consecutive channel uses is required. The conventional assumption is that two consecutive channel uses are approximately equal, i.e., $h_0[k]\approx h_0[k-1]$ and $h[k]\approx h[k-1]$. However, such an assumption is not valid for fast-fading channels.

For time-varying channels, individual channels are expressed by an AR(1) model as
\begin{gather}
\label{eq:ARi}
h_i[k]=\alpha_i h_i[k-1]+\sqrt{1-\alpha_i^2} e_i[k],\quad i=0,1, 2
\end{gather}
where $\alpha=\varphi_i(1)/\sigma_i^2$ is the auto-correlation of the $i$th channel and $e_i[k]\sim \mathcal{CN}(0,\sigma_i^2)$ is independent of $h_i[k-1]$. Based on these expressions, a first-order time-series model was derived in \cite{DAF-ITVT} to characterize the evolution of the cascaded channel in time. The time-series model of the cascaded channel is given as (see \cite{DAF-ITVT} for the detailed derivation and verification):
\begin{equation}
\label{eq:ARmodel}
h[k]=\alpha h[k-1]+\sqrt{1-\alpha^2}\ h_2[k-1]e_1[k],
\end{equation}
where $\alpha=\alpha_1 \alpha_2 \leq 1$ is the equivalent auto-correlation of the cascaded channel, which is equal to the product of the auto-correlations of individual channels, and $e_1[k]\sim \mathcal{CN}(0,\sigma_1^2)$ is independent of $h[k-1]$.

By substituting \eqref{eq:s-source}, \eqref{eq:ARi} and \eqref{eq:ARmodel} into \eqref{eq:y0} and \eqref{eq:y}, one has
\begin{equation}
\label{eq:y0k}
y_0[k]=\alpha_0 v[k] y_0[k-1]+\widetilde{w}_0[k],\\
\end{equation}
where
\begin{multline}
\label{eq:n0}
\widetilde{w}_0[k]=w_0[k]- \alpha_0 v[k] w_0[k-1]\\
+ \sqrt{1-\alpha_0^2} \sqrt{P_0} s[k]e_0[k]
\end{multline}
and
\begin{equation}
\label{eq:yk}
y_2[k]=\alpha v[k] y_2[k-1]+\widetilde{w}[k],
\end{equation}
where
\begin{multline}
\label{eq:ni}
\widetilde{w}[k]=w[k]- \alpha v[k] w[k-1]\\
+ \sqrt{1-\alpha^2} A\sqrt{P_0}h_2[k-1]s[k]e_1[k].
\end{multline}
It should be pointed out that, compared with slow-fading channels (see \cite[Eqs. (7) and 9]{DAF-SC}), additional terms appear in the noise expressions, which are functions of the channel auto-correlations and transmit power.

Then by substituting \eqref{eq:y0k} and \eqref{eq:yk} into \eqref{eq:zeta_0} and \eqref{eq:zeta_1}, one has
\begin{align}
\label{eq:z0_simp1}
\zeta_0= \Re \left\lbrace \alpha_0 v[k] |y_0[k-1]|^2+y_0^*[k-1]\widetilde{w}_0[k] \right\rbrace, \\
\label{eq:z1_simp1}
\zeta_2=\Re \left\lbrace \alpha v[k] |y_2[k-1]|^2+y_2^*[k-1]\widetilde{w}[k]\right\rbrace.
\end{align}
It is seen that, for given $y_0[k-1]$, $\zeta_0$ is a combination of complex Gaussian random variables, whose  conditional mean and variance are computed as 
\begin{multline}
\label{eq:mu_z0}
\mu_{\zeta_0}=\Et \{ \zeta_0 | y_0[k-1],v[k]=+1\}\\= \alpha_0 |y_0[k-1]|^2
+\Et\{\Re\{y_0^*[k-1] \widetilde{w}_0[k] \}\}\\=
\alpha_0 |y_0[k-1]|^2 -\alpha_0 \Et\{\Re\{w_0[k-1]|y_0[k-1]\} \} \\
= \alpha_0 |y_0[k-1]|^2 -\frac{\alpha_0 }{\rho_0+1} |y_0[k-1]|^2\\= \frac{\alpha_0 \rho_0}{\rho_0+1} |y_0[k-1]|^2,
\end{multline}
\begin{multline}
\label{eq:var_z0}
\Sigma_{\zeta_0}=\Var\{\zeta_0|y_0[k-1],v[k]=+1\}\\=
\Var\{\Re\{ \alpha_0 |y_0[k-1]|^2+y_0^*[k-1] \widetilde{w}_0[k] \} \}\\
= \Var\{\Re\{ y_0^*[k-1] \widetilde{w}_0[k] \} \}\\=
\frac{1}{2} \left[ N_0 + \alpha_0^2 \Var\{w[k-1]|y_0[k-1]\}  \right. \\ \left. +
(1-\alpha_0^2) P_0 \sigma_0^2
 \right] |y_0[k-1]|^2\\=
\frac{1}{2}N_0 \left( 1+ \frac{\alpha_0^2\rho_0}{\rho_0+1} +
(1-\alpha_0^2) \rho_0
 \right)|y_0[k-1]|^2.
\end{multline}

Furthermore, for given $y_2[k-1]$ and $h_2[k-1]$, $\zeta_2$ is a combination of complex Gaussian random variables and hence it is Gaussian as well. Its conditional mean and variance are computed as 
\begin{multline}
\label{eq:mu_z1}
\mu_{\zeta_2}=\Et \{ \zeta_2 | y_2[k-1],h_2[k-1],v[k]=+1\}\\= \alpha |y_2[k-1]|^2
+\Et\{\Re\{y_2^*[k-1] \widetilde{w}[k]|y_2[k-1],h_2[k-1] \}\}\\=
\alpha |y_2[k-1]|^2 -\alpha \Et\{\Re\{w[k-1]|y_2[k-1],h_2[k-1] \}\} \\
= \alpha |y_2[k-1]|^2 -\frac{\alpha }{\rho_2+1} |y_2[k-1]|^2\\= \frac{\alpha \rho_2}{\rho_2+1} |y_2[k-1]|^2,
\end{multline}
\begin{multline}
\label{eq:var_z1}
\Sigma_{\zeta_2}=\Var\{\zeta_2|y_2[k-1],h_2[k-1],v[k]=1\}\\=
\Var\{\Re\{ \alpha |y_2[k-1]|^2+y_2^*[k-1] \widetilde{w}[k] |y_2[k-1],h_2[k-1] \}\}\\
= \Var\{\Re\{ y_2^*[k-1] \widetilde{w}[k] |y_2[k-1],h_2[k-1] \}\}\\=
\frac{1}{2} \left( \sigma_w^2 + \alpha^2 \Var\{w[k-1]|y_2[k-1],h_2[k-1]\}\right.\\\left.  +(1-\alpha^2) P_0 \sigma_1^2 |h_2[k-1]|^2 \right) |y_2[k-1]|^2 =\\
\frac{1}{2}\sigma_w^2 \left[ 1+ \frac{\alpha^2\rho_2}{\rho_2+1} +
(1-\alpha^2) \rho_2 \right]|y_2[k-1]|^2.
\end{multline}

In the remaining of the paper, the time index $[k-1]$ is omitted for notational simplicity. From \eqref{eq:mu_z0} and \eqref{eq:var_z0}, the conditional pdf of $\zeta_0$ is given as
\begin{equation}
\label{eq:fz0_y0}
f_{\zeta_0}(\beta|y_0)=\frac{1}{\sqrt{2\pi \Sigma_{\zeta_0}}} \exp \left( - \frac{(\beta-\mu_{\zeta_0})^2}{2\Sigma_{\zeta_0}}\right).
\end{equation}
Since $y_0\sim \CN(0,N_0(\rho_0+1))$, $|y_0|^2\sim 0.5{N_0(\rho_0+1)}\chi_2^2$, i.e.,
\begin{equation}
\label{eq:f_y0}
f_{|y_0|^2}(\eta)=\frac{1}{N_0(\rho_0+1)} \exp\left( -\frac{\eta}{N_0(\rho_0+1)} \right).
\end{equation}
By taking the expectation of \eqref{eq:fz0_y0} over the distribution of $|y_0|^2$, the pdf of $\zeta_0$ is obtained as \cite[Eq. 3.471.15]{integral-tables} 
\begin{equation}
\label{eq:f_z0}
f_{\zeta_0}(\beta)=
\begin{cases}
b_0 \exp\left(c_0 \beta\right) & , \;\; \beta\leq 0 \\
b_0 \exp\left(d_0 \beta\right) & , \;\; \beta\geq 0,
\end{cases}
\end{equation}
where
\begin{gather}
\nonumber
b_0=\frac{1}{N_0(1+\rho_0)}, \\
\label{eq:b0c0d0}
c_0=\frac{2}{N_0(1+(1-\alpha_0)\rho_0)}, \\\nonumber
d_0=\frac{-2}{N_0(1+(1+\alpha_0)\rho_0)}.
\end{gather}

Thus, the cdf of $\zeta_0$ is expressed as
\begin{equation}
\label{eq:F_z0}
F_{\zeta_0}(\beta)=
\begin{cases}
\frac{b_0}{c_0} \exp\left(c_0 \beta\right) & , \;\; \beta\leq 0\\
1+\frac{b_0}{d_0} \exp\left(d_0 \beta\right) & , \;\; \beta\geq 0
\end{cases}.
\end{equation}
By substituting \eqref{eq:F_z0} into \eqref{eq:F_|z0|}, the cdf of $|\zeta_0|$ is obtained as
\begin{equation}
\label{eq:F_|z0|_sim1}
F_{|\zeta_0|}(\beta)=1+\frac{b_0}{d_0} \exp\left(d_0 \beta\right)- \frac{b_0}{c_0} \exp\left(-c_0 \beta\right).
\end{equation}

On the other hand, it follows from \eqref{eq:mu_z1} and \eqref{eq:var_z1} that the conditional pdf of $\zeta_2$ is
\begin{equation}
\label{eq:fz1_y2_h2}
f_{\zeta_2}(\beta|y_2,h_2)=\frac{1}{\sqrt{2\pi \Sigma_{\zeta_2}}} \exp \left( - \frac{(\beta-\mu_{\zeta_2})^2}{2\Sigma_{\zeta_2}}\right).
\end{equation}
Since, conditioned on $h_2$, $y_2\sim \CN(0,\sigma_{w}^2(\rho_2+1))$. Therefore, $|y_2|^2\sim 0.5{\sigma_w^2 (\rho_2+1)}\chi_2^2$, i.e.,
\begin{equation}
\label{eq:f_|y|}
f_{|y_2|^2}(\eta|h_2)=\frac{1}{\sigma_w^2(\rho_2+1)} \exp\left( -\frac{\eta}{\sigma_w^2(\rho_2+1)} \right).
\end{equation}
And by taking the expectation of \eqref{eq:fz1_y2_h2} over the distribution of $|y_2|^2$, one has \cite[Eq. 3.471.15]{integral-tables}
\begin{equation}
\label{eq:f_z1_h2}
f_{\zeta_2}(\beta|h_2)=
\begin{cases}
b_2 \exp\left(c_2 \beta\right) & , \;\; \beta\leq 0 \\
b_2 \exp\left(d_2 \beta\right) & , \;\; \beta\geq 0
\end{cases}
\end{equation}
where
\begin{gather}
\nonumber
b_2=\frac{1}{\sigma_w^2(\rho_2+1)}, \\
\label{eq:b2c2d2}
c_2=\frac{2}{\sigma_w^2(1+(1-\alpha)\rho_2)}, \\ \nonumber
d_2=\frac{-2}{\sigma_w^2(1+(1+\alpha)\rho_2)}.
\end{gather}
are functions of random variable $\lambda=|h_2|^2$, whose pdf is $f_{\lambda}(\lambda)=(1/\sigma_2^2)\exp(\lambda/\sigma_2^2)$.

Thus, the cdf of $\zeta_2$ conditioned on $h_2$ is given as
\begin{equation}
\label{eq:F_z1_h2}
F_{\zeta_2}(\beta|h_2)=
\begin{cases}
\frac{b_2}{c_2} \exp\left(c_2 \beta\right) & , \;\; \beta\leq 0\\
1+\frac{b_2}{d_2} \exp\left(d_2 \beta\right) & , \;\; \beta\geq 0
\end{cases}.
\end{equation}
By substituting \eqref{eq:F_z1_h2} into \eqref{eq:F_z1}, the cdf of $|\zeta_2|$, conditioned on $h_2$, is
\begin{equation}
\label{eq:F_|z1|}
F_{|\zeta_2|}(\beta|h_2)=1+\frac{b_2}{d_2} \exp\left(d_2 \beta\right)- \frac{b_2}{c_2} \exp\left(-c_2 \beta\right).
\end{equation}
Using \eqref{eq:f_z0} and \eqref{eq:F_|z1|}, \eqref{eq:Pb1} can be evaluated as follows:
\begin{multline}
\label{eq:Pb1_sim1}
P_{\bt}(E_1|h_2)= \int \limits_{-\infty}^{0} f_{\zeta_0}(\beta) \left[ F_{|\zeta_2|}(-\beta|h_2)-F_{|\zeta_2|}(0|h_2) \right] \dd \beta \\
=\int \limits_{-\infty}^{0} b_0 {\mathrm{e}}^{c_0 \beta} \left( \frac{b_2}{d_2} {\mathrm{e}}^{-d_2 \beta}-\frac{b_2}{c_2}{\mathrm{e}}^{c_2 \beta} -\frac{b_2}{d_2} +\frac{b_2}{c_2} \right) \dd \beta \\
= \frac{b_0b_2}{c_0} \left( \frac{1}{c_0-d_2} + \frac{1}{c_0+c_2}\right).
\end{multline}

Also, using \eqref{eq:F_|z0|_sim1} and \eqref{eq:f_z1_h2}, \eqref{eq:Pb2} is computed as
\begin{multline}
\label{eq:Pb2_sim1}
P_{\bt}(E_2|h_2)= \int \limits_{-\infty}^{0} f_{\zeta_2}(\beta|h_2) \left[ F_{|\zeta_0|}(-\beta)-F_{|\zeta_0|}(0) \right] \dd \beta \\
=\int \limits_{-\infty}^{0} b_2 {\mathrm{e}}^{c_2\beta} \left( \frac{b_0}{d_0} {\mathrm{e}}^{-d_0 \beta}-\frac{b_0}{c_0}{\mathrm{e}}^{c_0 \beta} -\frac{b_0}{d_0} +\frac{b_0}{c_0} \right) \dd \beta \\
= \frac{b_0b_2}{c_2} \left( \frac{1}{c_2-d_0} + \frac{1}{c_0+c_2}\right).
\end{multline}

Therefore, from \eqref{eq:Pb1_sim1} and \eqref{eq:Pb2_sim1}, the conditional BER can be expressed as
\begin{multline}
\label{eq:Pb_sim1}
P_{\bt}(E|h_2)=P_{\bt}(E_1|h_2)+P_{\bt}(E_2|h_2)\\=
\frac{b_0 b_2}{c_0c_2}+ \frac{b_0b_2}{c_0(c_0-d_2)}+\frac{b_0 b_2}{c_2(c_2-d_0)}.
\end{multline}

Finally, by substituting $b_2,c_2,d_2$ from \eqref{eq:b2c2d2} and taking the average over the distribution of $\lambda=|h_2|^2$, one has
\begin{equation}
\label{eq:Pb_sim2}
P_{\bt}(E)= I_1+I_2+I_3,
\end{equation}
where the terms $I_1$, $I_2$ and $I_3$ are determined in the following.

First, $I_1$ is computed as
\begin{multline}
\label{eq:I1}
I_1=\frac{b_0}{c_0} \int \limits_{0}^{\infty} \frac{b_2}{c_2}f_{\lambda}(\lambda) \dd \lambda=\frac{b_0}{c_0} \int \limits_{0}^{\infty} \frac{B_2}{2B_1}\frac{\lambda+B_1}{\lambda+B_2}\frac{1}{\sigma_2^2} {\mathrm{e}}^{-\frac{\lambda}{\sigma_2^2}} \dd \lambda\\=
\frac{b_0B_2}{2c_0B_1} \left(1+\frac{B_1-B_2}{\sigma_2^2}\exp\left(\frac{B_2}{\sigma_2^2}\right)E_1\left(\frac{B_2}{\sigma_2^2}\right) \right),
\end{multline}
where $B_1,\; B_2$ are defined as
\begin{equation}
\label{eq:B1B2}
\begin{split}
B_1=&\frac{1}{A^2\left(1+(1-\alpha)\rho_1\right)}, \\
B_2=&\frac{1}{A^2(1+\rho_1)}.
\end{split}
\end{equation}
Second, $I_2$ is obtained as
\begin{multline}
\label{eq:I2}
I_2=\frac{b_0}{c_0} \int \limits_{0}^{\infty} \frac{b_2}{c_0-d_2}f_{\lambda}(\lambda) \dd \lambda\\
=\frac{b_0}{c_0} \int \limits_{0}^{\infty} \tilde{B}_3\frac{1}{\lambda+\tilde{B}_2}\frac{1}{\sigma_2^2} {\mathrm{e}}^{-\frac{\lambda}{\sigma_2^2}} \dd \lambda\\=
\frac{b_0}{c_0}\tilde{B}_3 \left(\frac{1}{\sigma_2^2}\exp\left(\frac{\tilde{B}_2}{\sigma_2^2}\right)E_1\left(\frac{\tilde{B}_2}{\sigma_2^2}\right) \right),
\end{multline}
where
\begin{gather}
\label{eq:B1B2B3t}
\tilde{B}_2= \frac{(3+\alpha+(1-\alpha_0)\rho_0)\rho_1+3+(1-\alpha_0)\rho_0}{A^2(1+(2+\alpha)\rho_1+(1+\alpha)\rho_1^2)},
\\ \nonumber
\tilde{B}_3= \frac{(1+(1+\alpha)\rho_1)(1+(1-\alpha_0)\rho_0)}{2A^2(1+(2+\alpha)\rho_1+(1+\alpha)\rho_1^2)}.
\end{gather}

Third, $I_3$ is determined as
\begin{multline}
\label{eq:I3}
I_3=b_0 \int \limits_{0}^{\infty} \frac{b_2}{c_2(c_2-d_0)}f_{\lambda}(\lambda) \dd \lambda\\
= \int \limits_{0}^{\infty} \breve{B}_3\frac{\lambda+\breve{B}_1}{\lambda+\breve{B}_2}\frac{1}{\sigma_2^2} {\mathrm{e}}^{-\frac{\lambda}{\sigma_2^2}} \dd \lambda\\=
\breve{B}_3 \left(1+\frac{\breve{B}_1-\breve{B}_2}{\sigma_2^2}\exp\left(\frac{\breve{B}_2}{\sigma_2^2}\right)E_1\left(\frac{\breve{B}_2}{\sigma_2^2}\right) \right),
\end{multline}
where
\begin{gather}
\nonumber
\breve{B}_1=\frac{2}{A^2(1+(1-\alpha)\rho_1)}, \\
\label{eq:B1bB2bB3b}
\breve{B}_2=\frac{(3-\alpha+(1+\alpha_0)\rho_0)\rho_1+3+(1+\alpha_0)\rho_0}{A^2(1+(2-\alpha)\rho_1)+(1-\alpha)\rho_1^2)}, \\
\nonumber
\breve{B}_3=\frac{(1+(1-\alpha)\rho_1)^2(1+(1+\alpha_0)\rho_0)}{4(1+\rho_0)(1+(2-\alpha)\rho_1+(1-\alpha)\rho_1^2)}.
\end{gather}

In summary, the obtained BER expression in \eqref{eq:Pb_sim2} gives the exact BER of the D-AF relaying system using DBPSK and selection combining in time-varying Rayleigh fading channels {with general channel variances}. For the special case of $\alpha_0=1,\;\alpha=1,\;\sigma_i^2=1,\;i=0,1,2,$ this expression yields the BER of the system considered in \cite{DAF-SC} for the case of slow-fading {and symmetric channels}. It should be mentioned that, although, the BER expression of \cite[Eq.23]{DAF-SC} looks different than \eqref{eq:Pb_sim2} in the special case, both expressions give the same results for the case of slow-fading {and symmetric channels}. However, the expression in \eqref{eq:Pb_sim2} only involves computing the exponential integral function, whereas the expression of \cite[Eq.23]{DAF-SC} was derived in the integral form which also involves the exponential integral function. {The exponential integral function can be easily computed numerically.}

It is also seen that the obtained BER expression depends on the channel auto-correlations. This dependence is the reason for performance degradation in fast-fading channels and the fact that the BER reaches an error floor at high signal-to-noise ratio. The error floor can be obtained as (see Appendix for the proof):
\begin{equation}
\label{eq:PEf}
\lim \limits_{P_0/N_0 \rightarrow \infty} P_{\bt}(E)=\bar{I}_1+\bar{I}_2+\bar{I}_3,
\end{equation}
where
\begin{equation}
\label{eq:I1b}
\bar{I}_1=\lim \limits_{P_0/N_0\rightarrow \infty} I_1=\frac{1}{4}(1-\alpha_0)(1-\alpha),
\end{equation}
\begin{equation}
\label{eq:I2b}
\bar{I}_2=\lim \limits_{P_0/N_0\rightarrow \infty} I_2=
\frac{1-\alpha_0}{2\sigma_2^2}\bar{\tilde{B}}_3
\exp\left(\frac{\bar{\tilde{B}}_2}{\sigma_2^2}\right)E_1\left(\frac{\bar{\tilde{B}}_2}{\sigma_2^2}\right),
\end{equation}
with
\begin{gather}
\label{eq:lim_B2t}
\bar{\tilde{B}}_2=\lpn \tilde{B}_2= \frac{q \sigma_0^2}{1-q} \frac{1-\alpha_0}{1+\alpha},  \\
\label{eq:lim_B3t}
\bar{\tilde{B}}_3=\lpn \tilde{B}_3=\frac{q(1-\alpha_0)\sigma_0^2}{2(1-q)},
\end{gather}
and
\begin{equation}
\label{eq:I3b}
\bar{I}_3=\lpn I_3= \bar{\breve{B}}_3 \left(1-\frac{\bar{\breve{B}}_2}{\sigma_2^2} \exp\left(\frac{\bar{\breve{B}}_2}{\sigma_2^2} \right) E_1\left(\frac{\bar{\breve{B}}_2}{\sigma_2^2} \right) \right),
\end{equation}
with
\begin{gather}
\label{eq:lim_B2b}
\bar{\breve{B}}_2=\lpn \breve{B}_2 =\frac{q \sigma_0^2}{1-q} \frac{1+\alpha_0}{1-\alpha},  \\
\label{eq:lim_B3b}
\bar{\breve{B}}_3=\lpn \breve{B}_3 =\frac{1}{4} (1-\alpha)(1+\alpha_0).
\end{gather}
The above expressions show that the error floor is a function of the second-order statistics of the channels (auto-correlation and variance) and the power amplification factor and it is independent of the (high) transmitted power.

{
\section{Extension to Multi-Relay Systems}
\label{sec:mult-relay}
It is known that the achieved diversity order of the SC method with one relay in slow-fading channels is two \cite{DAF-SC}. In slow-fading channels, the cooperative diversity order can be increased by using more relays in the network. In a multi-relay system more channel information needs to be provided at the destination for the semi-MRC method to work. Therefore, using the SC method in multi-relay cooperative networks becomes even more attractive, as long as the trade-off between performance and complexity is satisfied. Similar to \eqref{eq:zeta_1}, the decision variables of all links can be computed and used in the selection combiner. Specifically, consider a multi-relay system with $L$ relays. The decision variables are computed as
\begin{equation}
\label{eq:zeta_2l}
\zeta_{2l}= \Re\{ y_{2l}^*[k-1] y_{2l}[k] \}, \quad l=1,\cdots,L
\end{equation}
where $y_{2l}$ is the received signal from the $l$th relay at the destination.
The output of the selection combiner for the multi-relay system is
\begin{equation}
\label{eq:zeta-sc-ml}
\zeta = \arg \max \limits_{\zeta_0,\{\zeta_{2l} \}_{l=1}^L}  \{ |\zeta_0|,|\zeta_{2}|,\cdots,|\zeta_{2L}| \}.
\end{equation}
Unfortunately, the theoretical BER analysis of the SC method in multi-relay system appears to be intractable. As such, only simulation results are presented in Section\ref{sec:sim} for multi-relay systems.
}

\section{Simulation Results}
\label{sec:sim}
In this section, first, the D-AF relay network {red}{with $L=1$ relay} is simulated for various channel qualities using both the SC and semi-MRC methods. The obtained theoretical BER and error floor of the SC method are verified with simulation results.

The channel coefficients are assumed to be Rayleigh flat-fading, i.e., $h_0[k]\sim \CN(0,\sigma_0^2),\; h_1[k]\sim \CN(0,\sigma_1^2),\; h_2[k]\sim \CN(0,\sigma_2^2).$ Based on the location of the nodes with respect to each other and channel qualities, variances of the channels would be different. Here, three scenarios are considered: (i) symmetric channels with $\sigma_0^2=1,\;\sigma_1^2=1,\; \sigma_2^2=1$, (ii) non-symmetric channels with strong SR channel $\sigma_0^2=1,\;\sigma_1^2=10,\;\sigma_2^2=1$, and (iii) non-symmetric channels with strong RD channel $\sigma_0^2=1,\;\sigma_1^2=1,\;\sigma_2^2=10$. The channel scenarios are summarized in Table~\ref{table:sig012}.
\begin{table}[!ht]
\begin{center}
\caption{Channel variances and corresponding optimum power allocation factors}
\label{table:sig012}
  \begin{tabular}{ |c | c| c|  }
    \hline
				& {$[\sigma_0^2,\sigma_1^2, \sigma_2^2]$}& $q_{\opt}$  \\ \hline\hline
{Symmetric}		& $[1,1,1]$    & 0.67   \\ \hline
{Strong SR } 		& $[1,10,1]$   & 0.58    \\ \hline
{Strong RD} 	& $[1,1,10]$   & 0.85    \\
    \hline
  \end{tabular}
\end{center}
\end{table}

The simulation method of \cite{ch-sim} is utilized to generate the time-correlated channel coefficients $h_0[k],h_1[k],h_2[k].$ The amount of time-correlation is determined by the normalized Doppler frequency of the underlying channel, which is a function of the mobility, carrier frequency and symbol duration. Obviously, for fixed carrier frequency and symbol duration, a higher mobility leads to a larger Doppler frequency and less time-correlation.

\begin{figure}[tb!]
\psfrag {f} [t][] [.9]{$f_0=f_1$}
\psfrag {BER} [][] [.9]{Error Floor}
\centerline{\epsfig{figure={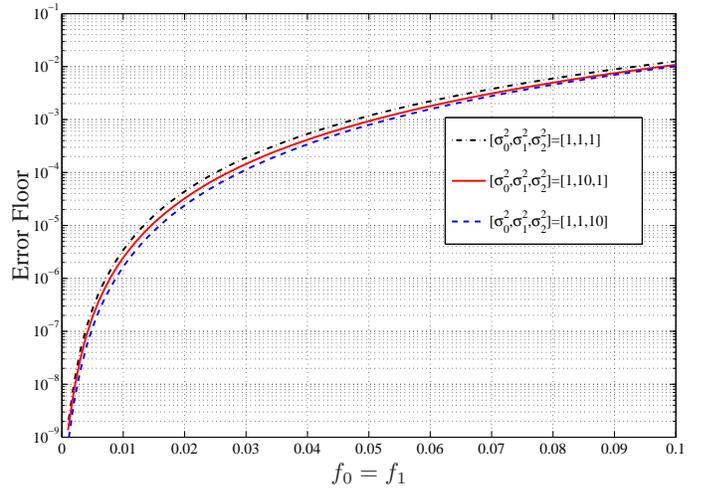},width=9cm}}
\caption{Error floors as functions of channel fade rates for different channel qualities, $f_0=f_1,\;f_2=0.001$.}
\label{fig:err_floor}
\end{figure}

Based on the normalized Doppler frequency of the three channels, different cases are considered. To get an understanding about choosing the normalized Doppler frequency values, the obtained error floor expression is examined for a large range of Doppler values. Fig.~\ref{fig:err_floor} plots of error floors versus channel fade rates for the three scenarios of Table~\ref{table:sig012}. It is assumed that the SD and SR channels have similar normalized Doppler frequencies, i.e., $f_0=f_1$, which changes from 0.001 to 0.1 and $f_2=0.001$. The plots in Fig.~\ref{fig:err_floor} can be divided into three regions. For fade rates less than 0.01, the error floor is very small and this region would be regarded as the slow-fading region. The fade rate of 0.01 is an approximate threshold beyond which the channels become relatively fast-fading. For fade rates between 0.01 to 0.05, the error floor increases asymptotically in a linear manner with a relatively sharp slope from $10^{-6}$ to $10^{-3}$. When the fade rate is larger than 0.05, the error floors continue to increase from $10^{-3}$ to $10^{-2}$. A BER value of around $10^{-2}$ is obtained in this region. This is very high for reliable communication and therefore the fade rate of 0.05 would be regarded as the threshold beyond which the channels become very fast-fading.

\begin{figure}[thb!]
\psfrag {time1} [][] [1]{$k$}
\psfrag {time2} [][] [1]{$n$}
\psfrag {h} [] [] [.8] {$|h[k]|$}
\psfrag {h0} [] [] [.8] {$|h_0[k]|$}
\psfrag {corr0} [] [] [.8] {$\varphi_0(n)$}
\psfrag {corr2} [] [] [.7] {$\varphi_1(n)\varphi_2(n)$}
\centerline{\epsfig{figure={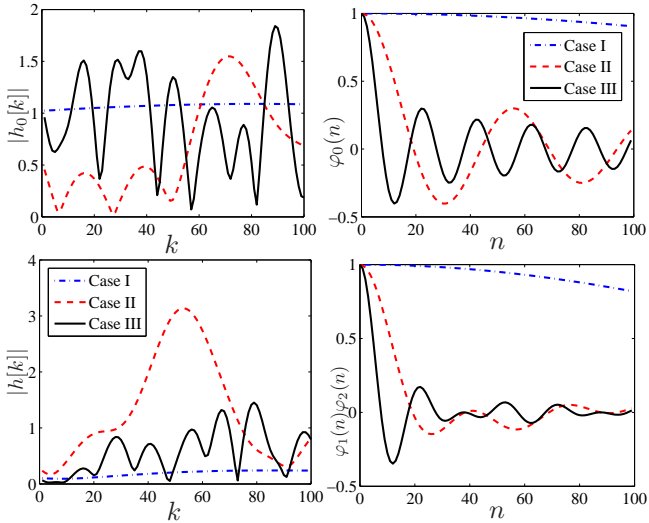},width=8.5cm}}
\caption{Snapshot of realizations of the direct and cascaded channels and the corresponding autocorrelations in different cases. Here $k$ and $n$ are defined as in Eq. \eqref{eq:phi}.}
\label{fig:doub-ch}
\end{figure}

From the discussion concerning Fig.~\ref{fig:err_floor}, three fading cases are considered. In Case I, it is assumed that all nodes are fixed or slowly moving so that all channels are slow-fading with the normalized Doppler values of $f_0=0.001,\;f_1=0.001,\;f_2=0.001$. In Case II, it is assumed that Source is moving so that the SD and SR channels are fast-fading with $f_0=0.02,\;f_1=0.02$ but Relay and Destination are fixed and the RD channel is slow-fading with $f_2=0.001$. In Case III, it is assumed that both Source and Destination are moving so that all the channels are fast-fading with $f_0=0.05,\;f_1=0.01$ and $f_2=0.05$, respectively. The normalized Doppler values are listed in Table \ref{table:f0f1f2}. Also, a snapshot of realizations of the direct and cascaded channels and their auto-correlation values in the three cases are plotted in Fig.~\ref{fig:doub-ch}. The plots show that when the normalized Doppler frequency values get larger, the channel coefficients fluctuate wider and the corresponding auto-correlation values decline faster over time. The auto-correlation of the cascaded channel declines faster than that of the direct channel as it involves the effects of two channels.
\begin{table}[!ht]
\begin{center}
\caption{Three fading cases.}
\label{table:f0f1f2}
  \begin{tabular}{ |c | c| c|c|c|  }
    \hline
				& $f_0$ & $f_1$ & $f_2$ & Channels status   \\ \hline\hline
{Case I }		& 0.001  & 0.001  & 0.001  & all channels are slow-fading \\ \hline
{Case II } 		& 0.02   & 0.02   &  0.001	& SD and SR are fast-fading \\ \hline
{Case III } 	& 0.05   & 0.01   &   0.05 & all channels are fast-fading \\
    \hline
  \end{tabular}
\end{center}
\end{table}

\begin{figure}[thb!]
\psfrag {BER} [] [] [1] {BER}
\psfrag {PAF} [t] [b] [1] {$q={P_0}/{P}$}
\psfrag {P/N_0} [t] [b] [1] {$P/N_0$}
\centerline{\epsfig{figure={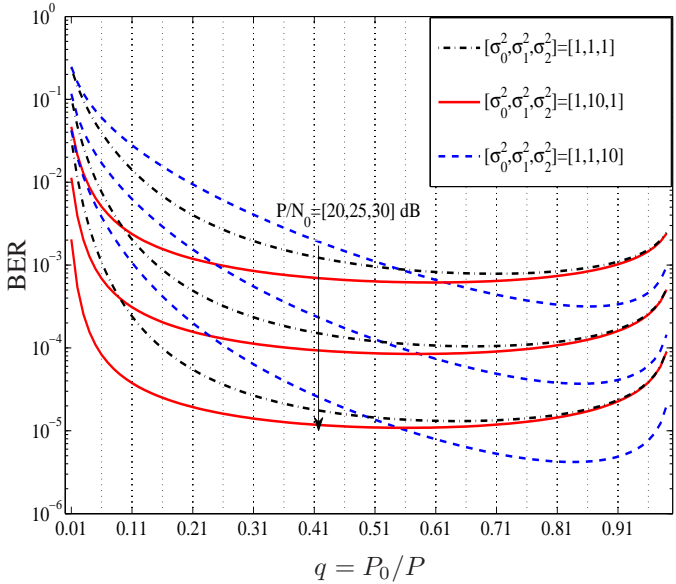},height=7.5cm,width=9cm}}
\caption{BER as a function of power allocation factor $q$ for $P/N_0=20,\;25,\;30$ dB.}
\label{fig:pw_all}
\end{figure}

First, the optimum power allocation between Source and Relay to minimize the BER is investigated in the three scenarios of Table~\ref{table:sig012} and fading rates of Case I (slow-fading). For each scenario listed in Table~\ref{table:sig012}, the BER expression is examined for different values of power allocation factor $q=P_0/P$, where $P=P_0+P_1$ is the total power. The BER curves are plotted versus $q$ in Fig.~\ref{fig:pw_all} for $P/N_0$=20,\;25,\;30 dB. Also, the optimum values obtained for the SC method are listed in Table~\ref{table:sig012} when $P/N_0=25$ dB. The optimum values in Table~\ref{table:sig012} and Fig.~\ref{fig:pw_all} show that in general more power should be allocated to Source than Relay. The BER is minimized at $q\approx 0.67$ and $\approx 0.58$ for symmetric and strong SR channels, respectively. When the RD channel becomes stronger than the SR channel, even more power should be allocated to Source and the BER is minimized at $q\approx 0.85$. This observation is similar to what reported in \cite[Table I]{DAF-MN-Himsoon} for the semi-MC technique.

\begin{figure}[th!]
\psfrag {P(dB)} [t][] [1]{$P/N_0$ (dB)}
\psfrag {BER} [] [] [1] {BER}
\psfrag {Lower Bound semi-MRC} [] [] [.9] {Lower Bound \cite{DAF-ITVT}}
\centerline{\epsfig{figure={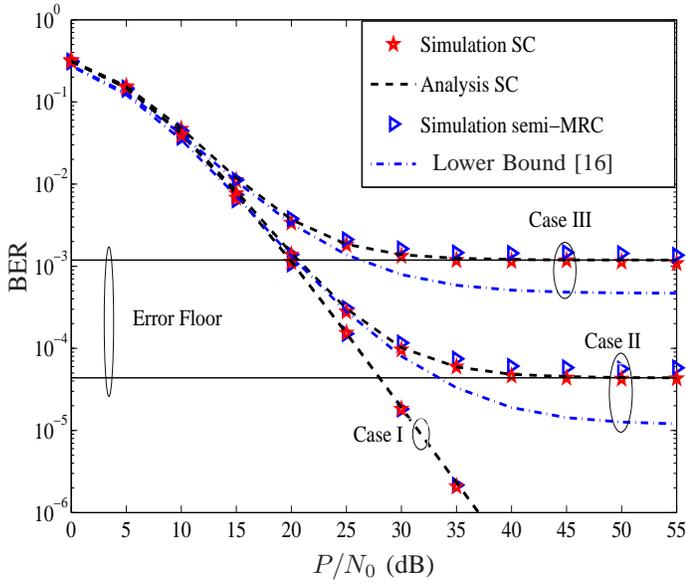},height=7.5cm,width=9cm}}
\caption{Theoretical and simulation BER of the D-AF system with SC and semi-MRC methods using DBPSK in different fading rates and symmetric channels: $\sigma_0^2=1,\sigma_1^2=1,\sigma_2^2=1$.}
\label{fig:scs_m2_sig11}
\end{figure}

\begin{figure}[hb!]
\psfrag {P(dB)} [t][] [1]{$P/N_0$ (dB)}
\psfrag {BER} [] [] [1] {BER}
\centerline{\epsfig{figure={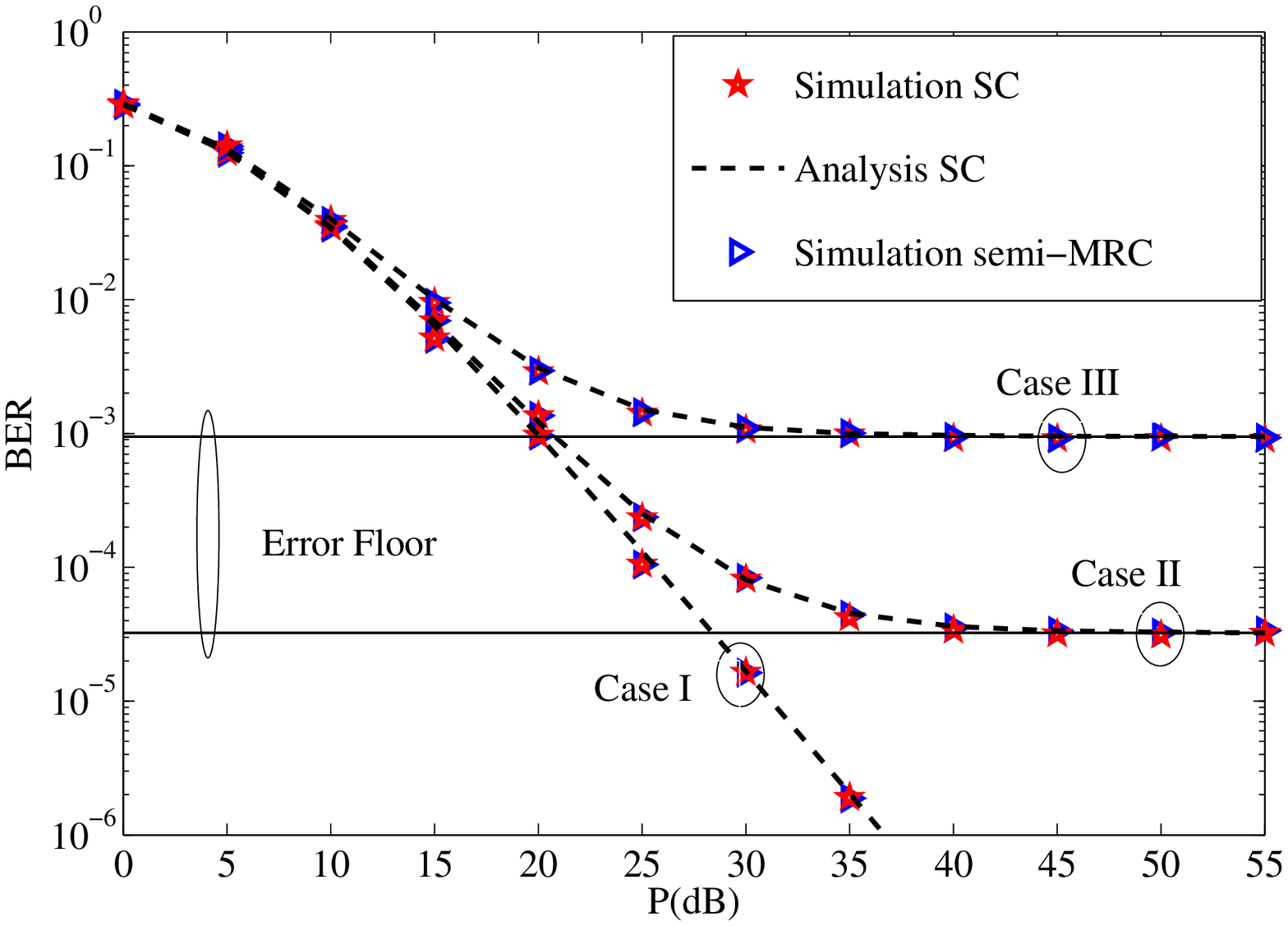},height=7.5cm,width=9cm}}
\caption{Theoretical and simulation BER of the D-AF system with SC and semi-MRC methods using DBPSK in different fading rates and strong SR channel: $\sigma_0^2=1,\sigma_1^2=10,\sigma_2^2=1$.}
\label{fig:scs_m2_sig101}
\end{figure}

\begin{figure}[hb!]
\psfrag {P(dB)} [t][] [1]{$P/N_0$ (dB)}
\psfrag {BER} [] [] [1] {BER}
\centerline{\epsfig{figure={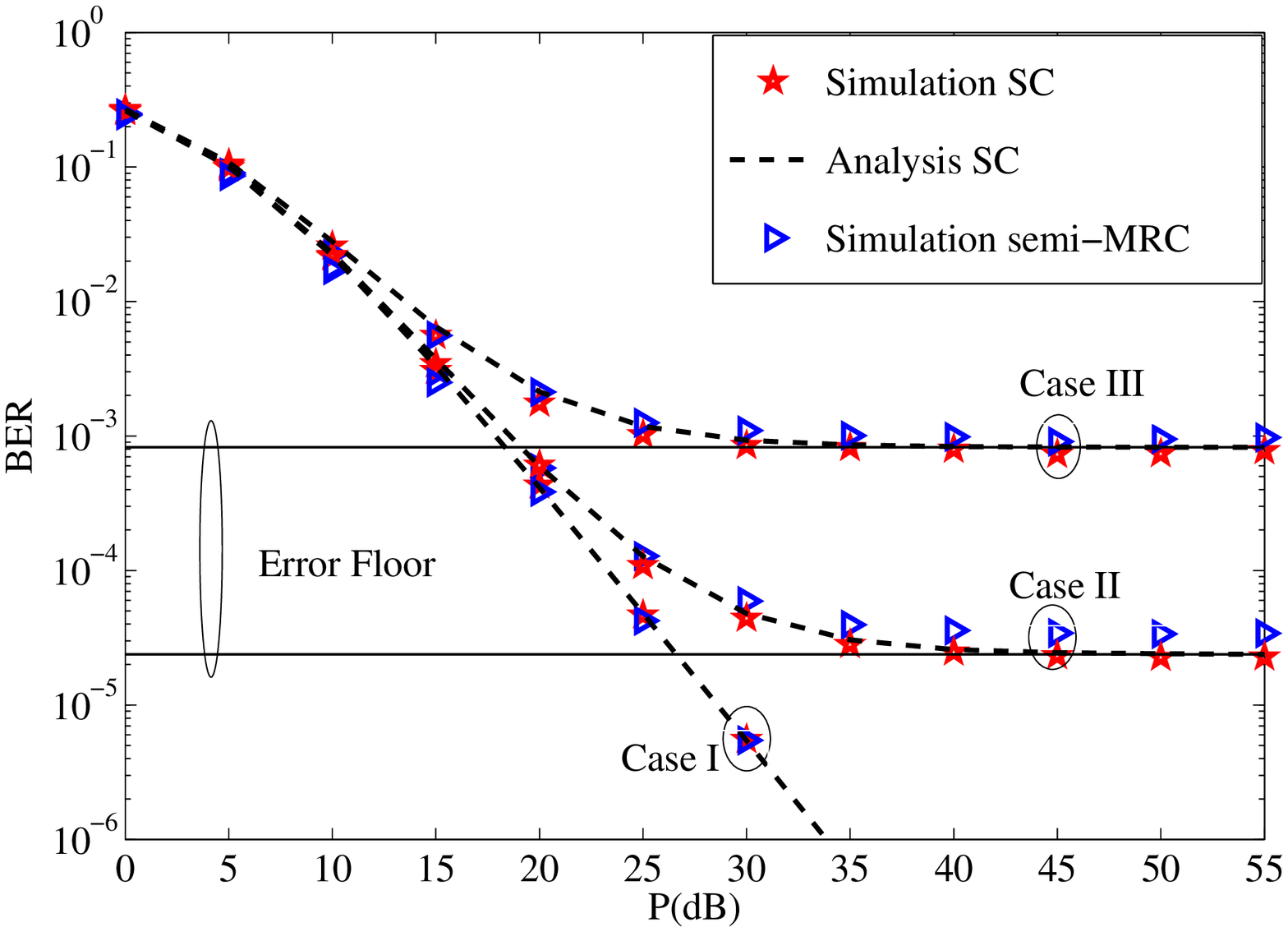},height=7.5cm,width=9cm}}
\caption{Theoretical and simulation BER of the D-AF system with SC and semi-MRC methods using DBPSK in different fading rates and strong RD channel: $\sigma_0^2=1,\sigma_2^2=1,\sigma_3^2=10$.}
\label{fig:scs_m2_sig110}
\end{figure}

The obtained power allocation factors are used in the simulation and the simulated BER using the SC and semi-MRC methods are computed for the three fading cases and channel variances. The BER results are plotted versus $P/N_0$ in Figs.~\ref{fig:scs_m2_sig11}--\ref{fig:scs_m2_sig110}. In each figure, three fading cases are shown. It should be mentioned that the results of Case I in Figure~\ref{fig:scs_m2_sig11} correspond to the results of \cite{DAF-SC} considered for the case of slow-fading {and symmetric channels}. On the other hand, the corresponding theoretical BER values of the SC method (for all cases) are computed from \eqref{eq:Pb_sim2} and plotted in Figs.~\ref{fig:scs_m2_sig11}--\ref{fig:scs_m2_sig110}. The horizontal lines show the theoretical values of the error floor for Case II and III, computed from \eqref{eq:PEf}. {For comparison purpose, the theoretical lower bound of the semi-MRC method as derived in \cite{DAF-ITVT} is also computed for symmetric channels and plotted in Fig.~\ref{fig:scs_m2_sig11}. It should be pointed out that the analysis of \cite{DAF-ITVT} only applies to symmetric channels.}

As can be seen from all the figures, the simulation results of the SC method are tight to the  analytical results for all fading cases and channel variances. This observation verifies our analysis. Specifically, in Case I (slow-fading) of all the figures, the BER of the SC method (and of the semi-MRC method as well) is consistently decreasing with $P/N_0$ and a diversity of two is achieved. The error floor in this case is very small ($\approx 10^{-9}$) and practically does not exist.

However, in Case II of all the figures, the situation is different from Case I. Since this case involves two fast-fading channels, the effect of channels variation is clearly observed in the obtained BER of both methods for $P/N_0>20$ dB. The BER gradually deviates from the results in Case I and eventually reaches to an error floor between $10^{-5}$ to $10^{-4}$. The exact amount of the error floors can be read from the horizontal lines in the figures.

Similarly, in Case III of all the figures, the obtained BER is degraded for $P/N_0>10$ dB. A severe degradation is seen in this case as all channels are fast-fading, specially two channels are around the threshold of very fast-fading region. There is no benefit in increasing the transmit power since an error floor around $10^{-3}$ appears for $P/N_0>30$ dB.

Moreover, the simulation results of both SC and semi-MRC methods are very close to each other in all the figures. In Cases II and III of Figs.~\ref{fig:scs_m2_sig11} and \ref{fig:scs_m2_sig110}, the results of the SC method are slightly better than that of the semi-MRC method at high $P/N_0$. This is due to the fact that the fixed combining weights (used in \eqref{eq:zeta}) of the semi-MRC method are not optimum and determined based on the second-order statistics of the channels and not their instantaneous CSI. Note that, in MRC method, the optimum combing  weights should be computed based on the noise variance of each link. The noise variance of the cascaded link is a function of the instantaneous CSI of the RD channel, which is not available in the considered system.

On the other hand, the close performance of both methods allows one to use the BER analysis of the SC method to tightly approximate the performance of the semi-MRC method in time-varying channels. It should be mentioned that the exact performance evaluation of the semi-MRC in time-varying channels appears difficult. {As a matter of fact, the lower bound reported in \cite{DAF-ITVT} is quite loose as can be seen in Fig.~\ref{fig:scs_m2_sig11}.}

\begin{figure}[hb!]
\psfrag {P(dB)} [t][] [1]{$P/N_0$ (dB)}
\psfrag {BER} [] [] [1] {BER}
\psfrag {L=2, Case III} [] [] [.9] {$L=2, $ Case III}
\psfrag {L=2, Case II} [] [] [.9] {$L=2, $ Case II \quad}
\psfrag {L=2, Case I} [] [] [.9] {$L=2, $ Case I\qquad}
\psfrag {L=3, Case I} [] [] [.9] {$L=3, $ Case I\qquad}
\psfrag {L=3, Case III} [] [] [.9] {$L=3, $ Case III}
\psfrag {simulation SC} [] [] [.9] {\qquad Simulation, SC}
\psfrag {simulation semi-MRC} [] [] [.9] {\qquad Simulation, semi-MRC}
\centerline{\epsfig{figure={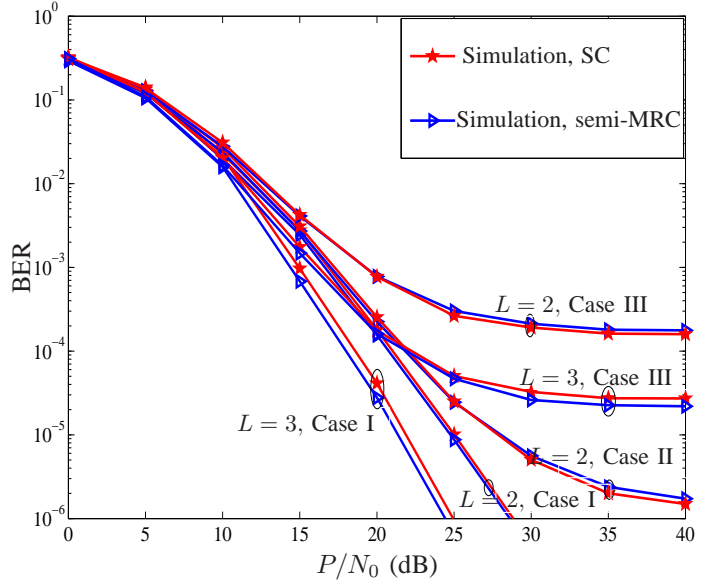},height=7.5cm,width=9cm}}
\caption{Simulation BER of D-AF systems with $L=2$ and 3 relays under different fading rates and symmetric channels.}
\label{fig:R2R3_scs}
\end{figure}

{
The diversity order can be improved by employing more relays. To see the effectiveness of the SC method in multi-relay networks, networks with $L=2$ and $L=3$ relays are also simulated. Symmetric channels and fading rates of Table~\ref{table:f0f1f2} are considered. In Fig.~\ref{fig:R2R3_scs}, the simulation results of the networks using both the SC and semi-MRC methods are plotted. It is clear that the diversity order is improved with increasing number of relays. More importantly, the difference between the BER performance of the SC and semi-MRC methods is negligible. This observation makes the SC method even more attractive for multi-relay networks since the semi-MRC method would require more channel information at the destination. In addition, the error floor is mitigated by employing more relays. 
}

\balance

\section{Conclusion}
\label{sec:con}
Selection combining of the received signals at Destination in a D-AF relay network employing DBPSK was studied in time-varying Rayleigh fading channels. Thanks to the differential encoding and selection combiner, no channel state information is needed at Destination for information detection. The exact bit-error-rate of the system was derived. Simulation results in various fading rates and channel variances verified the analysis and show that the selection combiner performs very close to the more-complicated semi-MRC method (which needs the second-order statistics of all channels at Destination). The analytical results show that the error performance depends on the fading rates of the equivalent channel and direct channel and an error floor exists at high signal-to-noise ratio region. The error floor was also analytically quantified.

\section*{Appendix}

\begin{center}
PROOF of \eqref{eq:PEf}
\end{center}

\begin{multline}
\label{eq:prf:I1b}
\bar{I}_1=\lim \limits_{P_0/N_0 \rightarrow \infty} I_1
= \lpn \frac{b_0B_2}{2c_0B_1}\\+
\lpn \frac{b_0B_2(B_1-B_2)}{2c_0B_1\sigma_2^2}\exp\left(\frac{B_2}{\sigma_2^2}\right)E_1\left(\frac{B_2}{\sigma_2^2}\right).
\end{multline}
From \eqref{eq:b0c0d0}, one has
\begin{equation}
\label{eq:b0/c0}
\lpn \frac{b_0}{c_0}=\lpn \frac{1+(1-\alpha_0)\rho_0}{2(1+\rho_0)}=\frac{1-\alpha_0}{2}.
\end{equation}
Likewise, taking the limit in \eqref{eq:B1B2} gives
\begin{multline}
\label{eq:B2/B1}
\lpn \frac{B_2}{2B_1}=\lpn \frac{A^2\left(1+(1-\alpha)\rho_1\right)}{2A^2\left(1+\rho_1\right)}\\=
\frac{1-\alpha}{2}.
\end{multline}
Again from \eqref{eq:B1B2}, one has
\begin{multline}
\label{eq:B1-B2}
B_1-B_2=\frac{1}{A^2\left(1+(1-\alpha)\rho_1\right)}-\frac{1}{A^2\left(1+\rho_1\right)}\\
= \frac{\rho_1}{\left(1+\rho_1\right)} \frac{\alpha}{A^2\left(1+(1-\alpha)\rho_1\right)}\\ \approx \frac{\alpha}{A^2\left(1+(1-\alpha)\rho_1\right)}=\alpha B_1,
\end{multline}
where the approximation has been made for large $P_0/N_0$.
Hence,
\begin{equation}
\label{eq:B2/B1(B1-B2)}
\frac{B_2(B_1-B_2)}{B_1}\approx \alpha B_2.
\end{equation}
On the other hand, for large $P_0/N_0$, $x=B_2/\sigma_2^2 \rightarrow 0$ and using the following approximation \cite[Eq. 5.1.20]{abro72ha}, one has
\begin{equation}
\label{eq:exE1x}
{\mathrm{e}}^x E_1(x)\approx \log \left(\frac{1}{x}\right)
\end{equation}
and
\begin{equation}
\label{eq:limexpE1x}
\lim \limits_{x \rightarrow 0} x {\mathrm{e}}^x E_1(x)=\lim \limits_{x \rightarrow 0} x \log(\frac{1}{x})=\lim \limits_{y \rightarrow \infty} \frac{\log(y)}{y}=0.
\end{equation}
Therefore, by substituting \eqref{eq:b0/c0}, \eqref{eq:B2/B1} and \eqref{eq:limexpE1x} into \eqref{eq:prf:I1b}, \eqref{eq:I1b} is obtained.

Finding $\bar{I}_2$ and $\bar{I}_3$ by taking the limit of $I_2$ and $I_3$ is straightforward. Note that, in deriving $\bar{I}_2$ and $\bar{I}_3$, by substituting $P_1=(1-q)P_0/q$ into \eqref{eq:A}, one has
\begin{equation}
\label{eq:A2q}
A^2=\frac{P_1}{P_0\sigma_1^2+N_0}=\frac{(1-q)P_0/N_0}{q(P_0/N_0\sigma_1^2+1)},
\end{equation}
\begin{equation}
\lpn \frac{1}{A^2}=\frac{q\sigma_1^2}{1-q}.
\end{equation}

\bibliographystyle{IEEEbib}
\bibliography{ref/references}

\end{document}